\documentclass[5p,times]{elsarticle}

\usepackage{graphicx}
\usepackage{amsmath}
\usepackage[noend]{algpseudocode}
\usepackage[linesnumbered,ruled,vlined]{algorithm2e}
\usepackage{multirow}
\usepackage{xcolor}

\usepackage{lscape}

\newcommand{\substepseparator}{\hspace{1cm}}
\usepackage{hyperref}

\journal{Int. J. Crit. Infrastructure Prot.}

\begin{document}

\begin{frontmatter}

\title{Adversarial Attacks and Mitigation for Anomaly Detectors of Cyber-Physical Systems\tnoteref{funding}}
\tnotetext[funding]{This work is partially supported by Singapore 
Economic Development Board and Fundamental Research Funds for the Zhejiang University NGICS Platform}

\author[1]{Yifan Jia\corref{cor1}}
\ead{yifan_jia@mymail.sutd.edu.sg}
\author[2]{Jingyi Wang}
\ead{wangjyee@zju.edu.cn}
\author[3]{Christopher M. Poskitt}
\ead{cposkitt@smu.edu.sg}
\author[1]{Sudipta Chattopadhyay}
\ead{sudipta_chattopadhyay@sutd.edu.sg}
\author[3]{Jun Sun}
\ead{junsun@smu.edu.sg}
\author[3]{Yuqi Chen}
\ead{yuqichen@smu.edu.sg}

\cortext[cor1]{Corresponding author}
\address[1]{Singapore University of Technology and Design, Singapore}
\address[2]{Zhejiang University, China}
\address[3]{Singapore Management University, Singapore}





\begin{abstract}
    The threats faced by cyber-physical systems~(CPSs) in critical infrastructure have motivated research into a multitude of attack detection mechanisms, including \emph{anomaly detectors} based on neural network models. The effectiveness of anomaly detectors can be assessed by subjecting them to test suites of attacks, but less consideration has been given to \emph{adversarial attackers} that craft noise specifically designed to deceive them. While successfully applied in domains such as images and audio, adversarial attacks are much harder to implement in CPSs due to the presence of other built-in defence mechanisms such as \emph{rule checkers} (or \emph{invariant checkers}). In this work, we present an adversarial attack that simultaneously evades the anomaly detectors and rule checkers of a CPS. Inspired by existing gradient-based approaches, our adversarial attack crafts noise over the sensor and actuator values, then uses a genetic algorithm to optimise the latter, ensuring that the neural network and the rule checking system are \emph{both} deceived. We implemented our approach for two real-world critical infrastructure testbeds, successfully reducing the classification accuracy of their detectors by over 50\% on average, while simultaneously avoiding detection by rule checkers. Finally, we explore whether these attacks can be mitigated by training the detectors on adversarial samples.
\end{abstract}


\end{frontmatter}


\section{Introduction} 

Cyber-physical systems~(CPSs), in which software components are deeply intertwined with physical processes, are ubiquitous in critical public infrastructure. The potential disruption that could result from a compromised system has motivated research into a multitude of CPS attack detection mechanisms, including techniques based on invariant checking~\cite{Adepu-Mathur16a,Feng-et_al19a,Yoong-Palleti-Silva-Poskitt20a,Yoong-et_al21a}, attestation~\cite{Valente-et_al14a,Roth-McMillin17a,Chen-Poskitt-Sun18a}, and fingerprinting~\cite{Formby-et_al16a,Ahmed-et_al20a}. A particularly popular solution is to build \emph{anomaly detectors}~\cite{Chandola:2009:ADS:1541880.1541882,Inoue-et_al17a,goh2017anomaly,fengLiChana,ZohrevandGlasserShahirTayebiCostanzo,Das-Adepu-Zhou20a,Schmidt-Hauer-Pretschner20a}, in which an underlying machine learning~(ML) model is trained on a time series of the system's physical data in order to judge when future values are deviating from the norm. Typically, this would be in the form of neural network, a model that is powerful enough to learn and recognise the complex patterns of behaviour that CPSs exhibit. Many studies have been performed in recent years to explore the efficacy of such deep learning anomaly detection approaches in CPSs~\cite{luo2020deep}.

The effectiveness of an anomaly detector can be assessed by subjecting it to a test suite of attacks, and observing whether it can correctly identify the anomalous behaviour. These tests can be derived from benchmarks~\cite{goh2016dataset}, hackathons~\cite{Adepu-Mathur18a}, or tools such as fuzzers~\cite{Chen-Poskitt-et_al19a,Chen-Xuan-Poskitt-et_al20a}, and typically involve manipulating or spoofing the network packets exchanged between CPS components. While studies have shown that neural network-based detectors are effective at detecting these conventional types of attacks~\cite{goh2016dataset,hodo2016threat,kosek_contextual_2016,sargolzaei_machine_2016,canizo2019multi,li2019deep,Kravchik-Shabtai18a}, less consideration has been given to testing their effectiveness at detecting \emph{adversarial attacks}, in which attackers have knowledge of the model itself and craft \emph{noise} (or \emph{perturbations}\footnote{We will use noise and perturbation interchangeably.}) that is specifically designed to cause data to be misclassified. If an anomaly detector fails to detect adversarial attacks, then the CPS relying on the detector is potentially at risk of a much broader range of attacks, since their effects can then simply be masked by this specially-crafted noise.

Adversarial attack algorithms have been applied across several different classification domains (including images~\cite{kurakin2016adversarial}, audio~\cite{carlini_audio_2018}, and malware~\cite{grosse2017adversarial}), but face a number of additional challenges to overcome in the context of CPSs in critical infrastructure. First, given that CPS anomaly detectors work by comparing the difference between actual and predicted system states, attackers can either attempt to enlarge this difference (promoting false positives) or shrink it (promoting false negatives). In other domains, adversarial attacks focus on the former, but for CPSs, the latter case---when a detector misclassifies a real attack as normal behaviour---can lead to serious consequences. Second, CPS states consist of both continuous and discrete sensor and actuator data, which leads to a complex interplay between the influence of noise and sensitivity to attacks. Finally, neural network-based detectors are rarely the only defence mechanism operating in real systems: typically, CPSs are also equipped with built-in \emph{rule checkers} that monitor for violations of known relationships between specific sensors and actuators (expressed in the form of \emph{invariants}). As a consequence, existing adversarial attacks are insufficient on their own as anomaly detectors and rule checkers must \emph{both} be deceived. Existing studies on evading neural network anomaly detectors (e.g.~\cite{Feng-et_al17a,Erba-et_al20a}) do not assume their presence.

In this work, we present an adversarial attack for testing recurrent neural network~(RNN) anomaly detectors of CPSs, which we assume to be equipped with additional rule checkers. Our approach is inspired by a white-box gradient-based approach~\cite{papernot2016limitations}, but adapted to meet the aforementioned challenges posed by critical infrastructure. In particular, our solution crafts noise over the continuous and discrete domains of sensors and actuators to deceive neural network-based anomaly detectors. Our experiments show that existing adversarial attacks have limited effectiveness in the presence of rule checkers, so we propose a genetic algorithm to optimise the combination of actuator values in order to \emph{simultaneously deceive} both the anomaly detector and the rule checkers.

We evaluate the effectiveness of our adversarial attacks by implementing them against RNN-based anomaly detectors for two real-world critical infrastructure testbeds: Secure Water Treatment (SWaT)~\cite{Mathur-Tippenhauer16a}, a multi-stage water purification plant; and Water Distribution (WADI)~\cite{ahmed_wadi:_2017}, a consumer distribution network. We demonstrate that our adversarial tests successfully and substantially reduce the accuracy of their anomaly detectors. Despite the general effectiveness of SWaT's built-in rule checkers, our genetic algorithm allows the attacker to evade them, while still achieving a similar accuracy reduction in the anomaly detector. Finally, we explore the possibility of mitigating attacks by training on adversarial samples, finding it is difficult in general to detect our attacks this way unless they involve a large amount of noise.

\substepseparator

\noindent\textbf{Contributions.} Our main contributions are summarised as follows:
\begin{itemize}

\item We define a threat model for CPS adversarial attackers, and propose a white-box gradient-based attack to construct adversarial tests for deceiving RNN-based anomaly detectors while simultaneously deceiving rule checkers.
\item We implement our adversarial attacker, and test it against RNN-based anomaly detectors of SWaT and WADI, two real-world critical infrastructure testbeds.
\item We find that our approach significantly reduces the accuracy of the detectors, i.e.~around 50\% on average, and evades detection by the rule checkers.

\end{itemize}

\section{Background}
\label{sec:background}

In this section, we state our assumptions about the structure of CPSs, and introduce two real-world examples that our work will be evaluated on. Following this, we define the threat model that our attacks will be based on, as well as what characterises a conventional and adversarial attack. Finally, we provide some background about the anomaly detectors we are testing, as well as the rule checkers that are assumed to be installed.

\subsection{Cyber-Physical Systems}
\label{sec:structure-of-a-cps}

In general, we assume that CPSs consist of two interconnected parts. First, a `physical' part, in which various physical processes are monitored by sensors and acted upon by actuators. Second, a `cyber' part, consisting of software components such as Programmable Logic Controllers~(PLCs)~\cite{alur_principles_2015} that implement some control logic. We assume that sensors, actuators, and PLCs are connected over a network, with PLCs taking sensor readings as input and computing commands to be returned to the actuators. Furthermore, we assume the presence of a Supervisory Control and Data Acquisition~(SCADA) system, connected to the PLCs, that can supervise the control process and issue commands of its own. 

We assume that CPSs contain two different types of countermeasures against attacks. First, the PLCs may contain some rule checkers, for checking that the sensor and actuator states pertain to some invariant properties (e.g.~if a certain sensor falls below a certain range, then a certain actuator should be open). Such rule checkers are standard for industrial CPSs. Second, we assume that a neural network-based anomaly detector~\cite{Giraldo-et_al18a} is present in the SCADA.

\substepseparator

\noindent\textbf{SWaT and WADI Testbeds.} We carry out our studies on two modern, real-world critical infrastructure testbeds: Secure Water Treatment (SWaT)~\cite{Mathur-Tippenhauer16a}, a water purification plant capable of producing five gallons of safe drinking water per minute; and Water Distribution~(WADI)~\cite{ahmed_wadi:_2017}, a consumer supply distribution system. The testbeds are funded by the Singapore Ministry of Defence and serve as key assets for research to be ported to protect actual public infrastructure plants. Though scaled-down, SWaT performs the same processes as actual water treatment plants, and was designed in close collaboration with engineers from PUB (Singapore's national water agency). In particular, SWaT is a six-stage plant including processes such as ultra-filtration, de-chlorination, and reverse osmosis. WADI consists of three stages, across which water is moved into elevated reservoirs, then supplied to consumer tanks on a pre-set pattern of demand.

\emph{Sensors and actuators.} In total, SWaT contains 68 sensors and actuators. Its sensors variously read the level of tanks, pressure, and flow across the system, whereas its actuators include motorised valves and pumps. Sensor readings are typically continuous values, whereas actuators are typically discrete (e.g.~`open' or `closed'; `on' or `off'). Note that a number of SWaT's actuators are `standbys' that are intended to be used only when the primary actuator fails. These are not considered in our work. After filtering such actuators, there are a total of 25 sensors and 26 actuators which are targeted in our study. WADI contains totally 127 sensors and actuators and we will use 70 sensors and 51 actuators after filtering, the sensor and actuators are the same types to those seen in SWaT.

The physical state of a testbed is a fixed ordering of all the sensor readings and actuator configurations at a particular timepoint. Table~\ref{tab:example} illustrates four such states, but for brevity, shows only the sensors and actuators involved in the first stage of SWaT (handling supply and storage). There are two sensors covering water flow (FIT101) and the tank level (LIT101), as well as a motorised valve (MV101) controlling at the inflow pipe of the tank, and a primary pump (P101) and secondary pump (P102) for pumping water out of the tank. The logged data over two seconds indicates that water is flowing into the tank and its level is increasing.

\begin{table}[t]
\centering
\caption{Example of log data from the SWaT historian}
\label{tab:example}
\footnotesize
\begin{tabular}{|l|l|l|l|l|l|l|}
\hline
\textbf{Time}              & \textbf{FIT101}   & \textbf{LIT101}   & \textbf{MV101} & \textbf{P101} & \textbf{P102} & \textbf{Status} \\ \hline
10:00:05 & 2.609 & 523.867 & 2     & 2    & 1    & Normal        \\ \hline
10:00:06 & 2.637 & 524.103 & 2     & 2    & 1    & Normal        \\ \hline
10:00:07 & 2.652 & 524.221 & 2     & 2    & 1    & Normal        \\ \hline
\end{tabular}
\end{table}

Formally, we will use $x$ to denote a system state, consisting of a fixed order of actuators and sensors: \[x = [x_{a1},x_{a2},x_{a3} \dots x_{s1},x_{s2},x_{s3} \dots]\]

\noindent Here, each $x_a$ represents an actuator value and each $x_s$ represents a sensor value. $X$ is the set of all possible system states. Let $S$ be a sequence of system states. We use $S[i]$ to denote the $i$-th state in the sequence, $S[i:j]$ to denote the sequence of system states from time $i$ to $j$, and $S[i:]$ to denote the sequence of states from time $i$ to the present.

\emph{Communication and control.} SCADA workstations are located in the plants' control rooms. A Human Machine Interface (HMI) is also located inside each plant and can be used to view the process states and set parameters. 
Control code can be loaded into the PLCs via the workstations. We can also acquire data logs (as well as network packet flows) from the historian at pre-set time intervals.

A multi-layer network enables communication across all components of SWaT and WADI. A ring network at Layer~0 of each stage enables the responsible PLC to communicate with sensors and actuators, whereas a star network at Layer~1 of the network enables communication between the PLCs. Supervisory systems such as the workstation and historian sit at levels further above.

\subsection{Threat Model} 
\label{sec:thread_model}

Given that we are testing the defence mechanisms of CPSs as an attacker, we define a threat model that states our assumptions of what an attacker knows about the system and is capable of doing within it.

As depicted in Figure~\ref{fig:CPSStructure}, our threat model assumes a white-box setting where an attacker is an insider that has the following capabilities:

\begin{figure*}[t]
  \centering
  \includegraphics[width=0.7\linewidth]{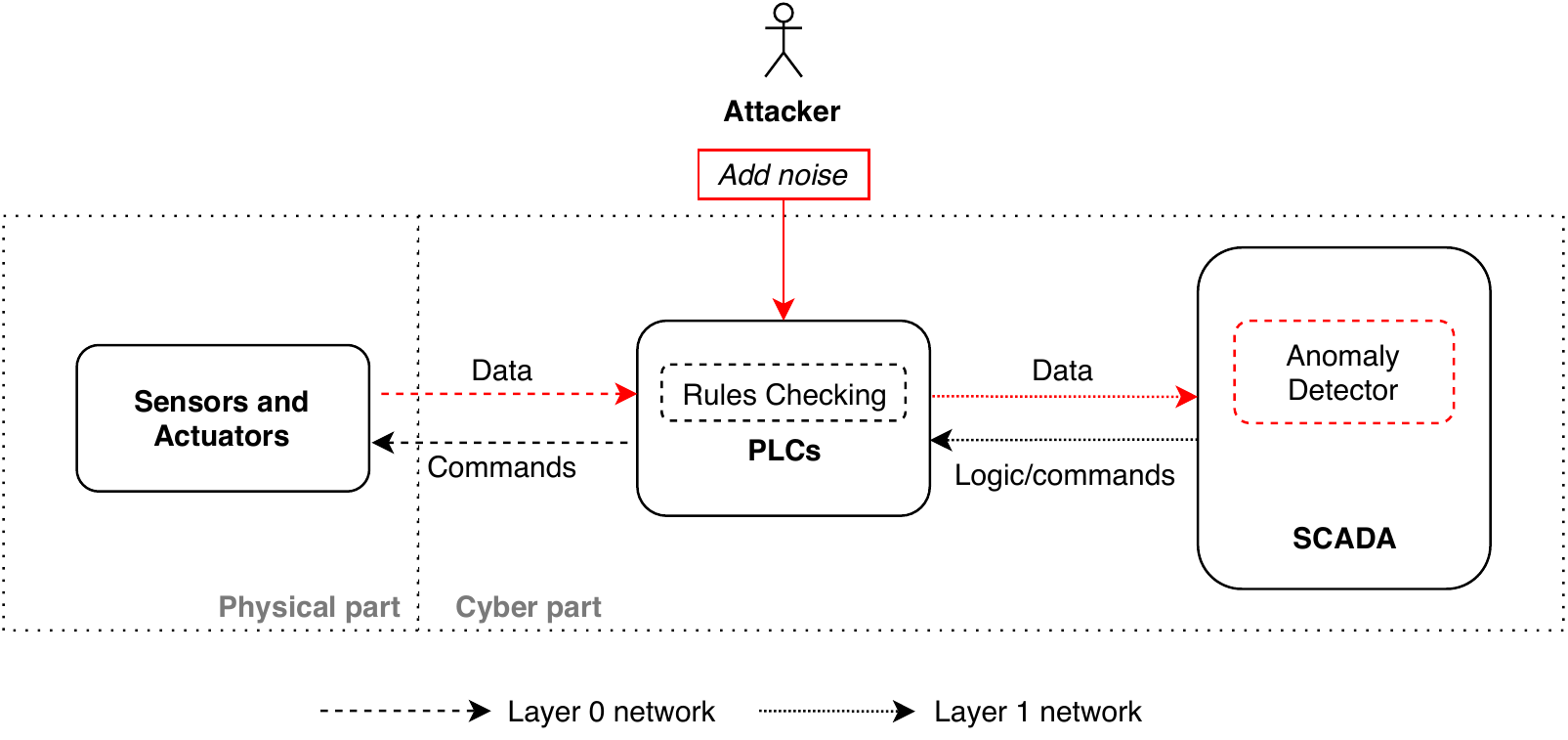}
  \caption{Overview of a cyber-physical system and an adversarial attacker}
  \label{fig:CPSStructure}
\end{figure*}

\begin{enumerate}
    \item The attacker is able to compromise the data transmitted from the physical part of the CPS to the PLCs at Layer~0 of the network.\label{cap1}
    \item The attacker has full knowledge about the RNN-based anomaly detector, including the network architecture, parameters, inputs, outputs, and other attributes.
    \item The attacker is aware of the presence of rule checkers, but cannot access any information about it other than inputs (i.e.~log data) and outputs (i.e.~`normal' or `anomalous'), which means attackers only know if the data can pass the rule or not.
\end{enumerate}
Compromising transmitted data as in capability (1) is a standard assumption in CPS attacks. We assume that our insider is able to do this via the connections between PLCs and sensors/actuators (i.e.~Level 0), but the results of this paper would also hold for the (easier to manipulate) star network at Level 1. Note that (2) is a standard assumption for adversarial attacks on neural networks in general (e.g.~\cite{papernot2016crafting}). We assume a more challenging black-box setting for rule checkers in (3), but our attacks can be applied in a grey- or white-box setting too. In particular, it would likely be easier to engineer a more powerful adversarial attack with grey- or white box knowledge of the rule checker systems.

\subsection{Attacks and Detection Mechanisms}
\label{sec:df_attacks}\label{tab:cps ad}

\noindent\textbf{CPS Attacks.} With the capabilities given by the threat model, \emph{conventional} CPS attacks are carried out by spoofing the sensor values that are transmitted from the physical part of the system to the PLCs, causing the control logic to issue the wrong actuator commands. For example, if a tank is near-empty, but an attacker spoofs a tank level sensor that is critically high, the pump could incorrectly be activated and lead to some underflow damage. Other examples are given in Table~\ref{tab:example3}. The SWaT and WADI testbeds have benchmarks~\cite{goh2016dataset} containing multiple different attacks of this kind, which have been used to test the effectiveness of different countermeasures~\cite{goh2016dataset,hodo2016threat,kosek_contextual_2016,sargolzaei_machine_2016}. Furthermore, data sets~\cite{goh2016dataset} are available containing several days of physical data resulting from subjecting the testbeds to these attacks. This data is suitable for training complex ML models such as RNNs and other kinds of neural networks.

\begin{table}[t]
\centering
\caption{Examples of manipulations}
\label{tab:example3}
\small
\begin{tabular}{|l|l|l|l|}
\hline
\textbf{Point} & \textbf{Start State} & \textbf{Attack} & \textbf{Intent} \\ \hline
MV‐101 & MV‐101 is closed & Open MV‐101 & Tank overflow \\ \hline
P‐102 & \begin{tabular}[c]{@{}l@{}}P‐101 is on \\ and P‐102 is off\end{tabular} & Turn on P‐102 & Pipe bursts \\ \hline
LIT‐101 & \begin{tabular}[c]{@{}l@{}}Water level \\ between L and H\end{tabular} & \begin{tabular}[c]{@{}l@{}}Increase by \\ 1 mm/s\end{tabular} & \begin{tabular}[c]{@{}l@{}}Tank Underflow;\\ Damage P‐101\end{tabular} \\ \hline
\end{tabular}
\end{table}

In this work, we aim to test CPSs against more than just these conventional attacks, by expanding the repertoire of attackers to include \emph{adversarial attacks}. Using their knowledge of the underlying RNN of the anomaly detector, adversarial attackers focus on crafting \emph{adversarial examples} that maximise some measure of harm, while masking their true effects from detectors by using carefully applied noise that deceive them.

To judge the success of a conventional attack, we can check whether at a certain time point $t$ there is an \emph{observable impact} in the physical state, $S[t]$, i.e.~the physical state differs from what it would have been in normal operation. As this is not simple to conclude in general, we leverage the operator-specified acceptable ranges of sensor values to identify that the system has been successfully attacked. For example, in SWaT, LIT101 indicates the water level in the first stage. If the reading is above 1100mm, then while it might not yet have overflowed, it is outside its acceptable range, and thus we conclude that the system is under attack. Note that in SWaT and WADI, these ranges are never entered during as part of normal operational behaviour---only in attack scenarios.

To judge whether an adversarial attack is successful is somewhat more complicated. Essentially, the goal is to deceive the anomaly detector and cause it to give an incorrect classification. For example, if the system is behaving normally and is being classified as such, the goal is to apply a minimal amount of noise such that the actual behaviour of the system does not change (i.e.~it remains normal), but the anomaly detector classifies it as anomalous (i.e.~a false alarm, decreasing confidence in the detector). On the other hand, if the attacker is spoofing sensor values and causing the behaviour to be classified as anomalous, the goal is to craft noise that does not affect the actual behaviour (i.e.~it remains anomalous in the same way) but causes the detector to classify it as normal. Thus, for the purpose of experimentation, it is important to be able to conclude that physical effects on the system \emph{before} the noise is applied are the same \emph{after} it is applied too.

\begin{figure}[t]
    \centering
    \includegraphics[width=1\linewidth]{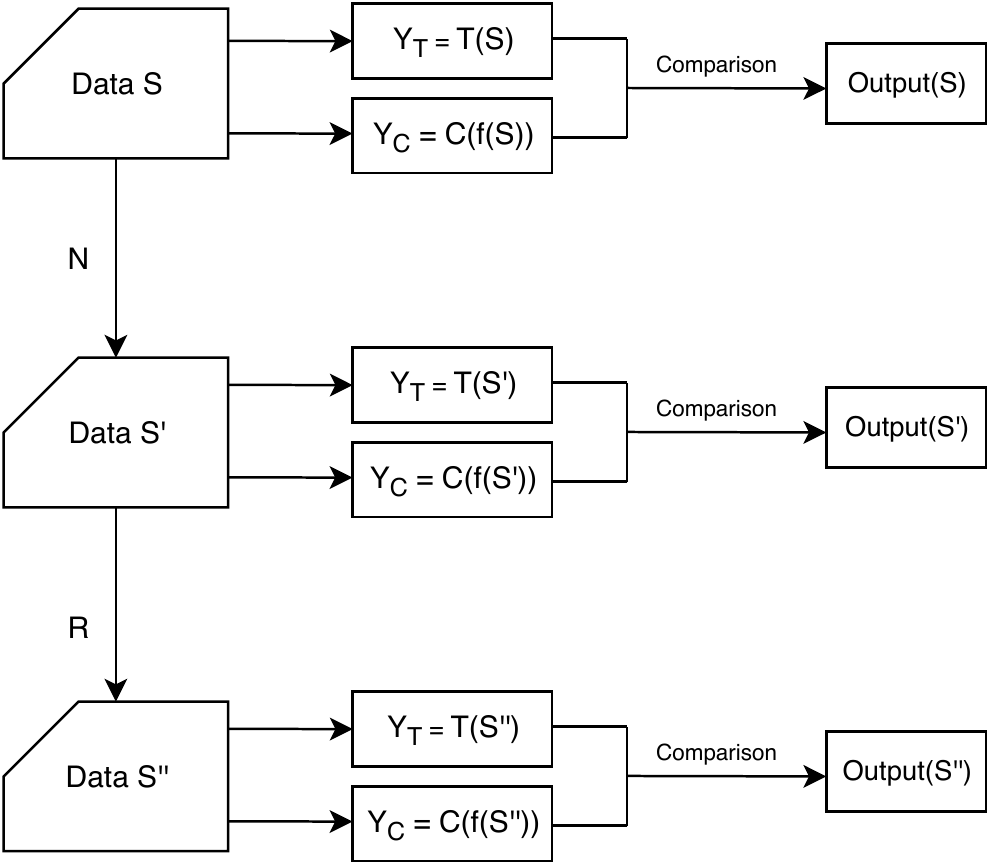}
    \caption{An overview of input vs output}
    \label{fig:gt}
\end{figure}

Since the definition of attacks is dynamic with respect to changes of input data, for every adversarial sample (normal or anomalous), we have to generate status labels as the ground truth. Figure~\ref{fig:gt} presents an overview of how results are generated. Vertically, Data S is the original attack data, Data S' is noisy data after a gradient-based attack by model N, and S'' is selected data by a genetic algorithm to pass rule checker R. (The details of these steps are given in Section~\ref{sec:method}.) Horizontally, from each data set (S, S' and S''), we calculate $Y_T$ using the ground truth function $T$ and $Y_C$ by CUSUM function $C$, where $f$ is the RNN predictor, and $C(f)$ represents the anomaly detector. We then get the $Output$ by comparing $Y_C$ and $Y_T$. The $Output$ inside the figure refers to a lists of standards such as precision, recall and f1. To illustrate this concretely, consider the original attack data S, and the corresponding data S' that results from the application of noise. Suppose a conventional attack is taking place (note the ground truth $Y_T$), and the anomaly detector correctly classifies this data as an attack ($Y_C$ = attack). In S', however, some noise has been applied to deceive the detector: the data is classified as normal ($Y_C$ = normal), but the actual physical effect on the system remains the same ($Y_T$ = attack). Intuitively, the original conventional attack is still taking place but has been masked by the noise. Note that if the level of noise is too much, or the wrong actuator is manipulated, it is possible to change the ground truth itself, i.e.~because the actual physical effect of the attack has changed.

\substepseparator

\noindent\textbf{Attack Detection Mechanisms.} Finally, we describe in more detail two different kinds of attack detection mechanisms, both in the context of SWaT: a rule checker~\cite{Adepu-Mathur16a} and an RNN-CUSUM anomaly detector~\cite{goh2017anomaly}.

\emph{Rule checking.} Adepu and Mathur~\cite{Adepu-Mathur16a} have systematically derived a set of invariants, consisting of 23 rules that describe the relationship between sensor and actuator values of SWaT. The idea of rule checking is to implement a set of pre-defined rules in the PLCs that should never be violated. For instance, a sensor value should never exceed its operation range, or regulations that the system should follow under normal operation. We use $R$ to denote the set of rules implemented in PLCs for checking. 
Once a rule $r\in R$ is violated, the system will raise an anomaly alarm to report that the system is in an abnormal state. Some rules are shown in Table~\ref{tab:example2}.
Let us take rule 1 as an example. The rule specifies that under the condition that sensor ``LIT101'' is equal or smaller than 500, the actuator ``MV101'' is supposed to be 2 (open) after 12 seconds.

\begin{table}[t]
\centering
\caption{Examples of rules}
\label{tab:example2}
\footnotesize
\begin{tabular}{|l|l|l|l|}
\hline
\textbf{Condition  }                                          & \textbf{Rule }             & \textbf{Time }\\ \hline
LIT101$\ \leq\ $500                                & MV101 = 2           & 12   \\ \hline
LIT101$\ \leq\ $250                                & P101 = 1 AND P102 = 1 & 2    \\ \hline
LIT301$\ \leq\ $800                                & P101 = 2 OR P102 = 2  & 12   \\ \hline
AIT201\ \textgreater{}\ 260 \& FIT201\ \textgreater{}\ 0.5 & P201 = 1 AND P202 = 1 & 2    \\ \hline
\end{tabular}
\end{table}

\emph{RNN-based anomaly detectors.} RNN-based anomaly detection has been adopted by many systems, especially systems generating time-series data~\cite{nanduri_anomaly_2016,filonov2017rnn,sheikhan2012intrusion,singh2017anomaly}. In both SWaT and WADI, the idea of such anomaly detectors is to predict the normal behaviour of the system based on historian data $S$ using a machine learning model (denoted by $f$). At run time, the system looks at the historian data (of a window-size length), uses $f$ to predict the system state $x'$, and compares it with the actual system state $x$. If the difference is beyond a threshold, the system is likely to be abnormal. In the work of Goh et al.~\cite{goh2017anomaly}, a RNN with LSTM architecture~\cite{Hochreiter:1997:LSM:1246443.1246450} is used as the prediction model, and a CUSUM algorithm (denoted by $C$) is used to calculate the differences between the actual value and the predicted value. This approach has been applied to the first stage of SWaT, and is shown to be effective for it. In our work, we re-implement their approach but for \emph{all} six stages of SWaT, and all three stages of WADI, as the anomaly detectors for us to test. As many systems have adopted RNN as a part of their anomaly detectors, our work is more general then simply SWaT/WADI.

\substepseparator

\noindent\textbf{Motivational Attack.} To help motivate the study that follows, we describe the steps that an attacker---an insider---could perform to successfully manipulate a CPS. First, using their knowledge of the detector, they could implement an adversarial sample generator on a device (e.g.~Raspberry Pi) installed between a PLC and its sensors/actuators. Raw data would pass through this generator first with a window size of 12 seconds, and the resulting noisy data would be optimised via a GA so as to pass the rule checkers. The generator would send this camouflaged noisy data to the PLC. The rule checkers and the RNN-based anomaly detectors that would assess this data are implemented inside the PLC and SCADA, i.e.~after the adversarial sample generator step. If the manipulated data avoids detection, then the consequences of the masked attack (e.g.~changing an actuator state) can then take effect on the system.

\section{Methodology and Design}
\label{sec:method}
Using the threat model and attack scenarios discussed in Section~\ref{sec:thread_model}, we aim to answer the following questions: (1) Is our adversarial attack effective on the anomaly detectors of real-world CPSs? (2) Can we design improved adversarial attack algorithms to deceive \emph{both} the anomaly detector and the rule checking system? We design experiments on actual critical infrastructure testbeds to answer these research questions. To decrease the performance of CPS anomaly detectors, we must overcome a number of challenges in our experiment design:

\begin{itemize}
    \item There are limited works on adversarial attacks on CPSs, especially for anomaly detectors with RNN models.
    \item There could be more than one defence mechanism inside the CPS (e.g. rule checkers). To complete an adversarial attack, we need to consider all defences.
    \item A CPS is typically complex and composed of multiple different sensors and actuators, each with different data types and ranges.
    \item CPSs are dynamic and hard to predict---certain changes may lead to the system shutting or breaking down. Statistical attacks in other domains do not face this problem. 
\end{itemize}

\begin{figure}
  \centering
  \includegraphics[width=1\linewidth,scale=0.5]{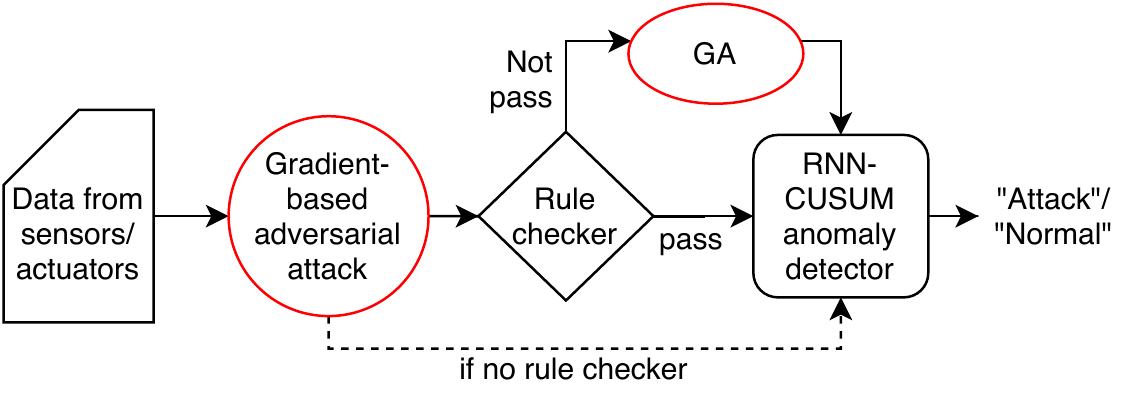}
  \caption{Workflow for adversarial attacks.}
  \label{fig:data_flow}
\end{figure}

We provide an an overview of our workflow in Figure~\ref{fig:data_flow}. For a basic CPS set up, data from sensors and actuators go directly to a RNN predictor for attack detection. We thus first apply a gradient-based adversarial attack to deceive the anomaly detector directly. As highlighted earlier, many CPSs also implement a rule checking system (see Section~\ref{tab:cps ad}) to check if the data respects some invariant properties. For such systems, if the adversarial attack is able to deceive a RNN predictor but is detected by the rule checking system, we further use a genetic algorithm in order to deceive the rule checker. Algorithm~\ref{algo:overall} shows the overall picture on how we generate adversarial examples to bypass both defences. In the following, we introduce the details step-by-step.

\begin{algorithm}[t]
\SetAlgoLined
\KwIn{Data $S_i$; RNN predictor model f; rule Checker R; noise level $\delta$}
\KwOut{adversarial examples}
Get gradient direction from f of loss w.r.t. inputs data $S_i$\;
Apply the noise $\delta$ to $S_i$ along or opposite the sign of gradients to get $S'_i$\;
Use Genetic Algorithm to generate $S_i''$ to pass rule checker\;
\Return{$S_i''$}
\caption{Overall Algorithm}
\label{algo:overall}
\end{algorithm}

\subsection{RNN predictor}
\label{sec:anomaly detection}

Following~\cite{goh2017anomaly}, we train an RNN in LSTM architecture from normal data $S_n$ as our prediction model.
The trained model is a many-to-one prediction model $f:S\to X$ which takes a certain sequence (parameterised by window size) of data historian as input and makes the prediction of the output of the coming timestamp. For example, a window size of 10 represents a sequential input of the past 10 timestamps.

Once we obtain the predicted output at each timestamp, we then adopt the CUSUM algorithm~\cite{bray_ordination_1957} to decide whether the system is in an abnormal state as follows. The difference $d$ at timestamp $i$ (denoted as $d[i]$) is calculated from the predicted value $x'[i]$ by the RNN model and the actual value $x[i]$; the difference $d$ will then be added cumulatively with an allowable slack $c$.
We calculate the CUSUM for each sensor with both positive and negative value by the following formula:
\begin{equation}
\begin{split}
    &SH[i] = max(0,SH[i-1] + d[i] - c)\\
    &SL[i] = min(0,SL[i-1] + d[i] + c)\\
    &d[i] = X'[i]  - x[i],
\end{split}
\end{equation}
where SH represents the set of high cumulative sum and SL represents the low cumulative sum, $d$ is the difference between the predicted value $x$ and actual value $x'$, and $c$ is the allowable slack which is defined as 0.05 multiplied by the standard deviation of $S$. Furthermore, two thresholds, i.e.~an Upper Control Limit (UCL) and a Lower Control Limit (LCL), for SH and SL to respectively compare with are required to check whether the system is in an abnormal state. Normally, UCL and LCL are defined according to an experiment for validating the training data. Table \ref{tab:UCL} shows the UCL and LCL from stage 1 of SWaT as an example.

\begin{table}[t]
\centering
\caption{Upper and Lower Control Limit}
\label{tab:UCL}
\begin{tabular}{| l| c| c| c| c|c| c| c|l|}
\hline
 {\bf Senor}& {\bf Upper Control Limit}&{\bf Lower Control Limit}\\
 \hline
 FIT101& 	2&	-1.5\\
  \hline
 LIT101&	50&	-2\\
 \hline
\end{tabular}
\end{table}

\begin{figure}[t]
    \centering
    \includegraphics[width=1\linewidth]{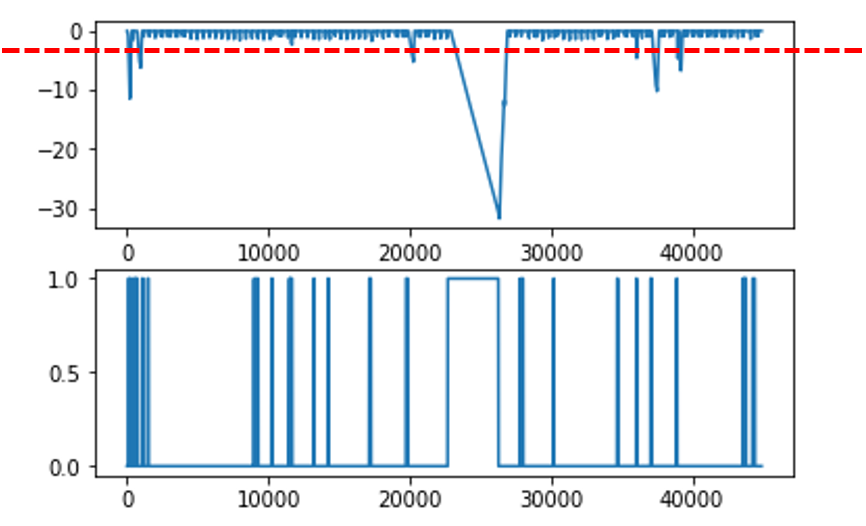}
    \caption{A CUSUM value example for sensor LIT101}
    \label{fig:CUSUM}
\end{figure}

The threshold selection for CUSUM is according to the validation of the attack data. We compare the CUSUM value for each sensor with attack labels, and set the threshold as the lowest peak of the CUSUM value that indicates an attack. Figure~\ref{fig:CUSUM} presents a CUSUM value example for sensor LIT101, the top graph is the negative CUSUM value $SL$ of LIT101. The bottom picture is the actual label of data: 1 means `under attack' and 0 means `normal status'. By comparing the two graphs, every peak value that is over the threshold from the $SL$ indicates an attack. We choose the lowest peak value as the threshold and it is indicated by the red dot line with value of $-2$. We calculate all thresholds accordingly.

To answer the research questions, we design experiments with the following concerns: RNN predictors (as described by Goh et al.~\cite{goh2017anomaly}) are trained using normal data, in which there are 51 features for SWaT and 21 features for WADI. Recall that these RNNs make a prediction of the data at time $t$ from a window of continuous data, which we use to compare with the actual records to decide if it is normal or an attack. In our experiments, we choose different window sizes for SWaT and WADI. For SWaT, we use 12s, which is the maximum delay from the rule checking system. For WADI, we use 10 timestamps, due to the slow speed of change in its physical processes. As the system is dynamic and the ground truth is changing whenever data is changing (see the difficulty discussed in Section~\ref{sec:method}), we compare the results of the ground truth model $Y_T = T(S)$ with the results from the anomaly detector $Y_C = C(f(S))$ to calculate the original anomaly detector accuracy of data S, and get the difference between $Y_T$ and $Y_C$ to get accuracy of data $S'$.

\subsection{Adversarial Attacks}
\label{sec:adversarial attack}

Our next step is to construct adversarial examples which aim to deceive the anomaly detectors. We consider two cases. Firstly, we construct adversarial examples to deceive only the RNN predictor. Secondly, we take a step further to deceive both the RNN predictor and the rule checking system, which is more realistic and challenging.  

In both cases, we assume a white-box attack in which the attacker can access the trained RNN model (and its parameters) and is able to compromise the data transmitted to the PLCs. The goal of the attacker is thus finding a minimum perturbation $\delta$ on the original input $\textbf{x}$ such that the detector will make a different decision from the original one. Formally, given the input $x$ at each timestamp and the RNN detector $f$, the objective of an attacker is to: 
\begin{equation}
\label{equ:cx}
\begin{split}
	& \min\ \delta\\
    & s.t.,\ f(\textbf{x}) \neq  f(\textbf{x}+\delta)\ and\ ||\delta|| \leq \tau
\end{split}
\end{equation}
where $\tau$ is a small value that restricts the manipulation range of $\textbf{x}$ according to a certain norm $||\cdot||$ (to measure the distance of the modification). The success of such an attack will deceive the detector in two ways. In the case that the detector detects an actual anomaly, the adversarial attack will be able to bypass the detector and put the system in danger (without noticing that the system is in the abnormal state). In the case that the detector detects an actual normal status, the adversarial attack will deceive the detector into raising false alarms. Though there are a lot of works on adversarial attacks on feed-forward networks in multiple domains~\cite{gong2017crafting,marra2018vulnerability,usama2018adversarial,huang2017adversarial}, \emph{there are relatively few on RNN as anomaly detectors and in the CPS domain which combines discrete and continuous values.} We introduce in detail how we solve the problem in the CPS setting.


\substepseparator

\noindent\textbf{Adversarial Attacks on RNNs.} In the primary setting, we aim to construct adversarial examples only for the RNN detector. 
In contrast to traditional adversarial attack domains for classification (e.g.~image or audio), the anomaly detector uses the CUSUM algorithm to check the difference between the predicted output and the actual output, then classifies normal or abnormal behaviour, with the system raising an anomaly alarm if the difference between them is beyond a threshold. Thus, we distinguish two scenarios in order to deceive the CUSUM checker.
The first scenario is to mask the normal states to be recognised as abnormal. This results in a similar case with Eq.~\ref{equ:cx} as follows. 

\begin{equation}
\label{equ:cx1}
\begin{split}
	& \min\ \delta\\
    & s.t.,\ ||f(\textbf{x}+\delta) -  y*)||\ge\Delta\ and\ ||\delta|| \leq \tau
\end{split}
\end{equation}
where $\textbf{x}$ is the original state, $y*$ is the actual output of the system, $\Delta$ is the acceptable difference (threshold for classifying anomaly). Notice that in this scenario the original state $\textbf{x}$ is classified as normal, i.e., $||f(\textbf{x}+\delta) -  y*)||<\Delta$. 

The second scenario is to mask an anomaly as a normal system state. The formalisation is slightly different as follows.  
\begin{equation}
\label{equ:cx2}
\begin{split}
	& \min\ \delta\\
    & s.t.,\ ||f(\textbf{x}+\delta) -  y*)||<\Delta\ and\ ||\delta|| \leq \tau
\end{split}
\end{equation}
Notice that in this scenario the original state $\textbf{x}$ is classified as anomalous, i.e., $||f(\textbf{x}+\delta) -  y*)||\ge\Delta$.

The remaining key challenge to solve Eq.~\ref{equ:cx1} and Eq.~\ref{equ:cx2} is to calculate the perturbation $\delta$. For this, we adopt the idea of gradient-based attacks~\cite{goodfellow_explaining_2014}, which perturbs the input \textit{along the gradient direction to maximise the change in the model prediction.}
The perturbation can be formalised as:
\begin{equation}\label{equ:del1}
\begin{split}
   & \textbf{x'} = \textbf{x} + \delta \\
    &\delta = \epsilon\cdot sgn(\nabla_x l(\textbf{x},y*))
\end{split}
\end{equation}
where $\textbf{x'}$ is the resultant adversarial example, $\epsilon$ is a constant representing the magnitude of the perturbation and $l$ is the loss function. In this work, we choose Mean Square Error (MSE) as the loss function. This is suitable for solving Eq.~\ref{equ:cx1} where we aim to find a perturbation which would enlarge the difference between $f(\textbf{x})$ and $y*$. On the other hand, to solve Eq.~\ref{equ:cx2}, we need to do the opposite and shrink the difference. To achieve this, we add perturbation along the \textit{opposite} direction of the gradient instead. That is,
\begin{equation}\label{equ:del2}
\begin{split}
   & \textbf{x'} = \textbf{x} - \delta \\
    &\delta = \epsilon\cdot sgn(\nabla_x l(\textbf{x},y*))
\end{split}
\end{equation}

Notice that in Eq.~\ref{equ:del1} and~\ref{equ:del2}, $\epsilon$ is a constant. This is not suitable for CPS setting with variables of discrete values. 
To solve the problem, we adapt the perturbation as follows and propose an adapted version of the Eq.\ref{equ:del1} and Eq.\ref{equ:del2} in practice:
\begin{equation}
\label{equ:del3}
\begin{split}
 &\textbf{x'} = 
    \begin{cases}
      	\textbf{x} + \delta ,  &if\ S_a = \text{normal} \\
        \textbf{x} - \delta,  &if\ S_a = \text{attack}\\
    \end{cases}\\
 &\delta = sgn(\nabla_\textbf{x}l(\textbf{x},y'))\cdot\Vec{\epsilon}
 \end{split}
\end{equation}

\noindent where $\Vec{\epsilon}$ is a diagonal matrix, and the value of each element from $\Vec{\epsilon}$ is defined according to the data type of each feature which represents the magnitude of the perturbation for each feature. The smaller that $\Vec{\epsilon}$ is, the less is the perturbation. 

We distinguish three different types of features (to solve the challenge introduced in Section~\ref{sec:method}), i.e.~actuators, valves and pumps, and thus add different perturbations on them. For sensors, we add a perturbation $\lambda$, which represents the percentage of the perturbation w.r.t. the original value. There are two types of actuators, valves and pumps. After normalisation, there are three values for valves: 0 for opening/closing, 0.5 for close and 1 for open. For pumps, there are two values: 0 for close and 1 for open. In order to follow the character of discrete values for actuators, we define the $\epsilon$ of valves as 0.5 and the $\epsilon$ of pumps is defined as 1. In summary, the diagonal matrix $\Vec{\epsilon}$ are defined as:
\[
\begin{split}
    &\Vec{\epsilon}_j{}_j =   
\begin{cases}
    \lambda, &if\ \Vec{F}[j]\in \Vec{U},\ where\ \Vec{U}\subset \Vec{F}\\
    0.5,  &if\ \Vec{F}[j]\in \Vec{V_m},\ where\ \Vec{V_m}\subset \Vec{F}\\
    1,  &if\ \Vec{F}[j]\in \Vec{V_p},\ where\ \Vec{V_p}\subset \Vec{F}\\
\end{cases}
\end{split}
\]
where $\Vec{F}$ is the set of all features, $\Vec{U}$ is the set of sensors, $\Vec{V_m}$ is the set of all valves, and $\Vec{V_p}$ is the set of all pumps. To give an example, if we only consider data from stage 1, $\Vec{\epsilon}=$
$$
\begin{vmatrix}
0.01&0&0&0&0\ \\
0&0.01&0&0&0\ \\
0&0&0.5\ &0&0\ \\
0&0&0&\ 1\ &0\ \\
0&0&0&0&\ 1\ \ \\
\end{vmatrix}
$$

In addition, to make sure the modified data are realistic, we always clip the adversarial examples to $[0,1]$ (with sensor values normalised). For example, 
for a pump with original value 1, if the sign of the gradient at $V_p[i,j]$ is positive, the perturbed pump value will be 2, which is not acceptable. We thus clip it back to 1.

To assess our research questions, we design an experiment to explore how much noise (i.e.~$\lambda$, the percentage of perturbation) is required for the attack to be effective at reducing the accuracy of the anomaly detector. We thus design the experiments as following: (1)~1\% noise ($\lambda = 0.01$) applied to all sensor values, but actuator values unchanged; (2)~10\% noise ($\lambda=0.1$) applied to all sensor values, but actuator values unchanged; (3)~1\% noise applied to all values, including actuators ($\lambda=0.5$ for valves and $\lambda=1$ for pumps); (4)~10\% noise applied to all values, including actuators ($\lambda=0.5$ for valves and $\lambda=1$ for pumps). The attack scenarios and parameters are enumerated in Table~\ref{tab:parameters} (GA parameters will be discussed in RQ2). The same logic has been applied to WADI.

\begin{table}[t]
\centering
\caption{Parameters for SWaT and WADI experiments}
\label{tab:parameters}
\begin{tabular}{|l|c|c|c|}
\hline
 & \textbf{sensor} & \multicolumn{2}{c|}{\textbf{actuator}} \\ \hline
  & \textbf{$\lambda$ (sensor)} & \textbf{$\lambda$ (valves)} & \textbf{$\lambda$ (pumps)} \\ \hline
\textbf{sensor (1\%)} & 0.01 & 0 & 0 \\ \hline
\textbf{sensor (10\%)} & 0.1 & 0 & 0 \\ \hline
\textbf{all (1\%)} & 0.01 & 0.5 & 1 \\ \hline
\textbf{all (10\%)} & 0.1 & 0.5 & 1 \\ \hline
\textbf{all (10\%) + GA} & 0.1 & 0.5 & 1 \\ \hline
\end{tabular}
\end{table}

Additionally, though the method we applied to RNN predictors is aiming to maximise the error of CUSUM, it is still able to be generalised to other neural networks with different targets. As the core idea is adding noise along gradients, which has been shown to be effective for different kinds of neural networks \cite{carlini_audio_2018,papernot2016crafting,gong2017crafting,goodfellow_explaining_2014}, the attacker can generate adversarial samples according to attack targets with other neural network models such as deep feed forward neural networks and convolutional neural networks. The attack targets could be wrong predictions, mis-classification, etc. 

In summary, we construct adversarial examples for the RNN detector by adding noise for two different attack scenarios in the above way. Algorithm~\ref{algo:gradient} shows the details. In case the constructed adversarial example could not pass the rule checking system at line 7, we move to the next step to bypass it further.

\begin{algorithm}[t]
\SetAlgoLined
\KwIn{Data $S_i$; RNN predictor model f}
\KwOut{adversarial examples}
\begin{math}
\delta \gets sgn(\nabla_\textbf{x}l(\textbf{x},y'))\cdot\Vec{\epsilon}\;
\end{math}\;
\eIf{$S_i$ is normal data}{
    $S_i'$ $\gets$ $S_i$+$\delta$\;
    }{
    $S_i'$ $\gets$ $S_i$-$\delta$\;
    }
\eIf{$S_i'$ cannot pass rule checker}{
    $S_i''$ $\gets$ GA(S'[i])\;
    }{
    $S_i''$ $\gets$ $S_i'$\;
    }
\Return{$S_i''$}
\caption{Adversarial examples construction for RNN detector}
\label{algo:gradient}
\end{algorithm}


\substepseparator

\noindent\textbf{Adversarial Attacks and Rule Checkers.} Rule checkers are widely used to detect anomalies of CPSs in many existing works \cite{Chen-Poskitt-Sun18a,alur_principles_2015,mitra_verifying_2013,akella_model-checking_2009}. In our experiment, we find that a gradient-based attack for RNN anomaly detectors alone could be easily detected by rule checkers in practice.
Therefore, we aim to construct adversarial examples to bypass both the RNN detector and the rule checking system as well by imposing an additional perturbation which is still as small as possible. 
Note that from our threat model, we assume the attackers only know if the data can pass the rule or not but do not have access to the logic behind them. It is thus infeasible to use constraint solvers to generate attacks satisfying these rules. 
To address this, our key observation is that the rule checking system makes decisions mainly depending on the status of actuators, i.e.~a minor perturbation on the sensor values will not influence the decisions of the rules. Thus, we propose to design a \emph{genetic algorithm (GA)}~\cite{goldberg_genetic_1989} to search for the ideal combination of actuator status to bypass the rule checking systems.


GAs~\cite{goldberg_genetic_1989} are a class of algorithms inspired by natural selection. In this work, we apply a GA to identify the ideal combination of actuator states which can bypass the rule checkers. Our GA consists of four main steps: (1)~Form an initial population of the actuator state combinations; (2)~Evaluate the fitness score for each candidate; (3)~Select candidates with highest fitness score as the ``parents''; (4)~Apply mutation or cross-over to generate ``children''. The algorithm will repeat step 2-4 until a candidate satisfies the termination condition or runs out of search budget. 

In our setting, a chromosome (a candidate solution) to form the population is the data point to manipulate ($S'[i]$). Since our objective is to select the best combination of actuator values, the sensor values need to be kept unchanged. However, the sensor values are taken into consideration since they are necessary to check if the rules are violated or not. As a result, both sensor and actuator values will be used to calculate the fitness score but only actuator values will be allowed for mutation or cross-over.
We define the fitness function considering two aspects: (1)~whether a candidate could pass the rules and (2)~how large is the modification. A candidate has a higher fitness score if it passes the rules with minimum modification. Formally, we define the fitness function as: 
\[g = c1*(1/||S'[i]-S[i]||)\]
The number of violated rules won't influence the results thus $c_1$ here is a Boolean variable indicating whether the candidate could pass the rule checker and $||S'[i]-S[i]||$ represents the modification level. Based on the fitness score, we assign each candidate a probability for selection by normalisation, i.e.~a candidate with higher score will have a higher probability to be chosen. In practice, we use roulette wheel selection \cite{goldberg_genetic_1989}.

After selection, we separate out the actuator values of the selected candidate for mutation with probability $p_m = 0.5$. 
For cross-over, we choose one-point cross-over to select only one random cross-over point to align with our purpose of minimising the modification.
Next, we calculate the fitness score for all candidates and select $n$ fittest candidates to form the new population. This procedure continues until the fittest candidate passes the rule checker or we run out of iterations.

\section{Evaluation on Cyber-Physical Systems}
\label{sec:application}

In this section, we conduct experiments to assess the effectiveness of our adversarial attacks on the anomaly detectors of real critical infrastructure testbeds, especially in the presence of rule checkers. We also investigate whether the anomaly detectors can be adapted to mitigate the threat of adversarial attacks.

\subsection{Preliminaries}
\label{subsec:swatDataset}

Our study is based on the SWaT dataset \cite{goh2016dataset} and WADI dataset \cite{ahmed_wadi:_2017}, which are publicly available \cite{swatDataset} and have been used in multiple projects \cite{goh2017anomaly,Chen-Poskitt-et_al19a,Chen-Xuan-Poskitt-et_al20a,wang2018towards,raman2019anomaly,Chen-Poskitt-Sun16a,shalyga2018anomaly}. The SWaT dataset records the system state of 26 sensor values and 25 actuators (in total 51 features) every second, and WADI records 70 sensors and 51 actuators (in total 121 features) every second. The sensor values are integer or floating-point numbers while the actuator values are discrete, e.g.~0 (opening/closing), 1 (closed), or 2 (open). The datasets consist of two types of data:

\begin{itemize}
\item \noindent{\em Normal data}: The normal dataset $S_n$ of SWaT was collected over seven days (a total of 496,800 records) when the system was under normal operation, and the dataset of WADI was collected for 14 days (a total of 172,800 records). The data is used to train the LSTM-RNN predictor.
\item \noindent{\em Attack data}: The attack dataset $S_a$ of SWaT was collected for four days (a total of 449,909 records) consisting of 36 attacks with labels (normal or attack),  and the dataset of WADI was collected for two days (a total of 172,801 records). The data is used as testing data for the anomaly detector. 
\end{itemize}

The neural network models are trained using the Keras~\cite{chollet2015} platform. All the experiments are conducted on a laptop with 1 Intel(R) Core(TM) i7-8750H CPU at 2.20GHz, 16GB system memory, and for the GPU, NVIDIA GeForce GTX 1050 Ti with Max-Q Design. Generating adversarial attacks for SWaT---4 days of data with 51 features---took 4-6 hours varied for different noise, while for WADI, it took 4-6 hours for 3 days of data with 121 features. The code and all the experiment results are available at~\cite{gitlink}.

\subsection{Experiments and Results}

Our experiments are designed to answer the following three research questions.

\paragraph{RQ1 (Are our adversarial attacks effective on anomaly detectors of real-world CPSs?)}

The results for different SWaT and WADI parameters are given in Table~\ref{tab:swat_attacks_rule}. We observe that the precision, f1, and accuracy have all been reduced after the adversarial attack. Intuitively, accuracy appears to decrease as more noise is added, both in terms of percentage and range of data it is applied to. However, it is possible that with the larger amounts of noise, the noisy data has exceeded the threshold defined as an attack by the model T. Thus the precision, recall and f1 for $10\%$ noise are all similar or higher than $1\%$ noise. Comparing to the original performance of the anomaly detector, the reduced accuracy has shown that the anomaly detector is vulnerable and sensitive to the adversarial attacks.

\begin{table*}[t]
\centering
\caption{Accuracy of the RNN-CUSUM anomaly detector for SWaT/WADI under different attack scenarios}
\label{tab:swat_attacks_rule}
\begin{tabular}{|l|cccc|cccc|}
\hline
\multicolumn{1}{|c|}{{\color[HTML]{000000} }} & \multicolumn{4}{c|}{{\color[HTML]{000000} \textbf{SWaT}}} & \multicolumn{4}{c|}{{\color[HTML]{000000} \textbf{WADI}}} \\ \hline
{\color[HTML]{000000} \textbf{Attack}} & \multicolumn{1}{c|}{{\color[HTML]{000000} \textbf{prec.}}} & \multicolumn{1}{c|}{{\color[HTML]{000000} \textbf{recall}}} & \multicolumn{1}{c|}{{\color[HTML]{000000} \textbf{f1}}} & {\color[HTML]{000000} \textbf{accuracy}} & \multicolumn{1}{c|}{{\color[HTML]{000000} \textbf{prec.}}} & \multicolumn{1}{c|}{{\color[HTML]{000000} \textbf{recall}}} & \multicolumn{1}{c|}{{\color[HTML]{000000} \textbf{f1}}} & {\color[HTML]{000000} \textbf{accuracy}} \\ \hline
{\color[HTML]{000000} \textbf{none}} & {\color[HTML]{000000} 0.96} & {\color[HTML]{000000} 0.86} & {\color[HTML]{000000} 0.91} & {\color[HTML]{000000} 89.40\%} & {\color[HTML]{000000} 0.62} & {\color[HTML]{000000} 0.70} & {\color[HTML]{000000} 0.66} & {\color[HTML]{000000} 71.36\%} \\ \hline
{\color[HTML]{000000} \textbf{sensor (1\%)}} & {\color[HTML]{000000} 0.62} & {\color[HTML]{000000} 0.53} & {\color[HTML]{000000} 0.57} & {\color[HTML]{000000} 52.38\%} & {\color[HTML]{000000} 0.43} & {\color[HTML]{000000} 0.65} & {\color[HTML]{000000} 0.52} & {\color[HTML]{000000} 52.85\%} \\ \hline
{\color[HTML]{000000} \textbf{sensor (10\%)}} & {\color[HTML]{000000} 0.65} & {\color[HTML]{000000} 0.58} & {\color[HTML]{000000} 0.61} & {\color[HTML]{000000} 48.78\%} & {\color[HTML]{000000} 0.42} & {\color[HTML]{000000} 0.67} & {\color[HTML]{000000} 0.52} & {\color[HTML]{000000} 49.88\%} \\ \hline
{\color[HTML]{000000} \textbf{all (1\%)}} & {\color[HTML]{000000} 0.52} & {\color[HTML]{000000} 0.5} & {\color[HTML]{000000} 0.51} & {\color[HTML]{000000} 51.85\%} & {\color[HTML]{000000} 0.34} & {\color[HTML]{000000} 0.96} & {\color[HTML]{000000} 0.5} & {\color[HTML]{000000} 33.28\%} \\ \hline
{\color[HTML]{000000} \textbf{all (10\%)}} & {\color[HTML]{000000} 0.12} & {\color[HTML]{000000} 0.84} & {\color[HTML]{000000} 0.21} & {\color[HTML]{000000} \textbf{28.84\%}} & {\color[HTML]{000000} 0.33} & {\color[HTML]{000000} 0.95} & {\color[HTML]{000000} 0.49} & {\color[HTML]{000000} \textbf{32.23\%}} \\  \hline
\end{tabular}
\end{table*}

The final row of the table, where noise is applied to both sensor and actuator values at the same time, has lower accuracy than the sensor-only parameters; this is because involving actuators results in a stronger impact on the RNN predictor, fed through to the CUSUM calculation. The results of precision and recall imply that many false alarms have been generated by the anomaly detector, indicating that changing actuators is easy to be caught by an anomaly detector as an attack. However, the recall of 0.84 in the SWaT table is high: this is because of fewer actuators were changed when we calculated the gradients (i.e.~most actuators have gradient 0).

We remark that for WADI, ground truth was determined in a slightly different way to SWaT: instead of using control points (which are not provided for WADI), we calculate the operation range under normal conditions with a small tolerance to set the ground truth model T. As it has more features (121) than SWaT (51), the complexity is more apparent and behaviour is more unpredictable. Therefore the anomaly detector has a lower accuracy on WADI compared to SWaT. However, the overall results still show the gradient-based noise could reduce the accuracy of the anomaly detector, and more noise has a better effect.

The results before and after adding the noise suggest our adversarial attack can reduce the accuracy of RNN-CUSUM anomaly detectors significantly. For the weakest attack of 1\% noise on sensors only, the accuracy and f1 (model performance) has been substantially reduced, and the more noise we add, the less accurate the anomaly detector is. However, applying noise to actuators sometimes can reduce the accuracy notably (10\% in SWaT), and sometimes insignificantly (1\% in SWaT). This observation indicates that successful adversarial attacks on CPSs need to consider the complex interaction between sensor and actuator values.  

\begin{center}
\noindent\fbox{%
    \parbox{0.75\linewidth}{%
        \small\emph{Our adversarial attacks are effective on RNN-based anomaly detectors of real-world CPSs, e.g.~reducing accuracy from 89.40\% to 28.84\% in SWaT.}
    }
}
\end{center}

\paragraph{RQ2 (How should the attacker compromise the original data to deceive both the rule checkers and the anomaly detector?)}

Rule checking systems are commonly used in industrial CPSs as a basic defence mechanism. As WADI has not yet established a rule checking system, our second experiment is only applied to the SWaT testbed, which monitors invariants related to a total of 20 actuators and 11 sensors. To overcome this, our experiment applies noise to all data, then fine-tunes the actuator states by using a GA in order to pass the rule-checker. We assume that the attacker does not have the access to the content of the rule, but knows whether the data could pass the rule checker or not. The rules include the relationship between the sensor and actuator values, so we divided a data point $x$ into $x_a = [x_{a1},x_{a2}\dots]$ and $x_s = [x_{s1},x_{s2}\dots]$, and use the GA to generate possible $x_a$ and keep $x_s$ to combine back to calculate the fitness score. As our goal is to generate adversarial attacks rather than break down the system, we only use our GA to replace attacks that are being detected by rule checker due to noise.

During the experiments, we found that 88.12\% of our adversarial attacks were detected by rule checkers, with the accuracy of anomaly detection for the attacks that evaded them increasing to 86.11\%, indicating rule checkers are very effective and necessary for CPSs. Moreover, we found there are some rules are more dominating than others, such as RULE19\footnote{If AIT402$<$240,  P403=1 AND P404=1 after 2 seconds}. Among detected attacks, 71.28\% break RULE19. We also found that seven of the rules were never violated.

The results of this experiment are given in Table~\ref{tab:swat_attacks_pass_rule_checkers}. From the table we can see that before we apply our GA, the rule checkers could catch most of our adversarial attacks and increase back the accuracy of detector to 86.11\%. However, attacks using a GA have successfully reduced the accuracy to 39.56\% without being detected by the rule checker. The accuracy is slightly higher compared to before (28.84\%), indicating that using a GA for actuator value selection may sacrifice a drop of accuracy which is reasonable, as GA generated values do not change following the sign of gradients. Besides this, overall performance is almost the same as the same parameter that does not make use of a GA, suggesting that this one allows us to pass the rule checker without much reducing the anomaly detector performance.

\begin{table}[t]
\centering
\caption{Accuracy of the RNN-CUSUM anomaly detector for SWaT during attacks that \emph{also} deceive the rule checkers}
\label{tab:swat_attacks_pass_rule_checkers}
\begin{tabular}{|l|cccc|cccc|}
\hline
{\color[HTML]{000000} \textbf{Attack scenarios}} & \multicolumn{1}{c|}{{\color[HTML]{000000} \textbf{precision}}} & \multicolumn{1}{c|}{{\color[HTML]{000000} \textbf{recall}}} & \multicolumn{1}{c|}{{\color[HTML]{000000} \textbf{f1}}} & {\color[HTML]{000000} \textbf{accuracy}}  \\ \hline
{\color[HTML]{000000} \textbf{all (10\%)}} & {\color[HTML]{000000} 0.91} & {\color[HTML]{000000} 0.93} & {\color[HTML]{000000} 0.92} & {\color[HTML]{000000} 86.11\%}  \\ \hline
{\color[HTML]{000000} \textbf{all (10\%) + GA}} & {\color[HTML]{000000} 0.14} & {\color[HTML]{000000} 0.76} & {\color[HTML]{000000} 0.24} & {\color[HTML]{000000} \textbf{39.56\%}}  \\ \hline
\end{tabular}
\end{table}

\begin{center}
\noindent\fbox{%
    \parbox{0.75\linewidth}{%
        \small\emph{The attacker can use a GA to optimise actuator values of an adversarial sample to deceive both rule checkers and the anomaly detector.}
    }
}
\end{center}

\paragraph{RQ3 (Can we design a defence against such adversarial examples?)}

The adversarial examples have more noise compared to original data, so we design a \emph{defence neural network (def-NN)} and a random forest~(RF) classification model to distinguish if the data is from an adversarial example or data from the system. If the noise is high enough, at least one of the classifiers should be able to detect the noisy data. The def-NN and RF perform as two binary classifiers. We train models using a dataset consisting of 50\% original data and 50\% adversarial examples, reserving a portion (with the same ratio) as testing data. The def-NN model is designed with three layers and 100 units for each layer, is trained for 6 minutes with 100 units, with $binary\ cross\ entropy$ as our loss function with a epoch of 20. The random forest model is implemented using the scikit-learn machine learning library~\cite{scikit-learn}, and is trained for 5 minutes with 10 estimators.

The results of SWaT and WADI can be found in Table~\ref{tab:swat_defence_nn}. The training data and testing data has been split with a ratio of 7:3. The table shows the f1 score and accuracy of testing data with different noise levels as predicted by models fed with different types of training data. As the noisy data and normal data have equal weights, the accuracy for the overfitting problem is 0.50. 

\begin{table*}[t]
\centering\footnotesize
\caption{def-NN and RF for detecting adversarial attacks on SWaT and WADI}
\label{tab:swat_defence_nn}
\begin{tabular}{|c|c|cccc||cccc|}
\hline
\multicolumn{2}{|c|}{} & \multicolumn{4}{c|}{\textbf{def-NN Testing data}} & \multicolumn{4}{c|}{\textbf{RF Testing data}} \\ \cline{3-10} 
\multicolumn{2}{|c|}{\multirow{-2}{*}{f1/accuracy}} & \multicolumn{1}{c|}{\textbf{sensor(1\%)}} & \multicolumn{1}{c|}{\textbf{sensor(10\%)}} & \multicolumn{1}{c|}{\textbf{all(1\%)}} & \textbf{all(10\%)} & \multicolumn{1}{c|}{\textbf{sensor(1\%)}} & \multicolumn{1}{c|}{\textbf{sensor(10\%)}} & \multicolumn{1}{c|}{\textbf{all(1\%)}} & \textbf{all(10\%)} \\ \hline
 & \textbf{sensor(1\%)} & \textbf{0.99/0.99} & 0.99/0.99 & 0.73/0.78 & 0.95/0.96 & \textbf{0.99/0.99} & 0.99/0.99 & 0.99/0.99 & 0.99/0.99 \\ \cline{2-10} 
 & \textbf{sensor(10\%)} & 0.00/0.50 & \textbf{0.99/0.99} & 0.15/0.54 & 0.44/0.64 & 0.00/0.50 & \textbf{0.99/0.99} & 0.01/0.5 & 0.99/0.99 \\ \cline{2-10} 
 & \textbf{all(1\%)} & 0.00/0.50 & 0.00/0.50 & \textbf{1.00/1.00} & 0.99/0.99 & 0.07/0.52 & 0.20/0.56 & \textbf{1.00/1.00} & 0.99/0.99 \\ \cline{2-10} 
\multirow{-4}{*}{\textbf{\begin{tabular}[c]{@{}c@{}}SWaT\\ Training\\ data\end{tabular}}} & \textbf{all(10\%)} & 0.00/0.50 & 0.00/0.50 & 1.00/1.00 & \textbf{1.00/1.00} & 0.01/0.50 & 0.94/0.94 & 0.98/0.98 & \textbf{1.00/1.00} \\ \hline\hline
{\color[HTML]{000000} } & {\color[HTML]{000000} \textbf{sensor(1\%)}} & {\color[HTML]{000000} \textbf{1.00/1.00}} & {\color[HTML]{000000} 1.00/1.00} & {\color[HTML]{000000} 1.00/1.00} & {\color[HTML]{000000} 1.00/1.00} & {\color[HTML]{000000} \textbf{1.00/1.00}} & {\color[HTML]{000000} 1.00/1.00} & {\color[HTML]{000000} 1.00/1.00} & {\color[HTML]{000000} 1.00/1.00} \\ \cline{2-10} 
{\color[HTML]{000000} } & {\color[HTML]{000000} \textbf{sensor(10\%)}} & {\color[HTML]{000000} 0.99/0.99} & {\color[HTML]{000000} \textbf{1.00/1.00}} & {\color[HTML]{000000} 0.99/0.99} & {\color[HTML]{000000} 1.00/1.00} & {\color[HTML]{000000} 0.99/0.99} & {\color[HTML]{000000} \textbf{1.00/1.00}} & {\color[HTML]{000000} 0.99/0.99} & {\color[HTML]{000000} 1.00/1.00} \\ \cline{2-10} 
{\color[HTML]{000000} } & {\color[HTML]{000000} \textbf{all(1\%)}} & {\color[HTML]{000000} 0.00/0.50} & {\color[HTML]{000000} 0.00/0.50} & {\color[HTML]{000000} \textbf{1.00/1.00}} & {\color[HTML]{000000} 1.00/1.00} & {\color[HTML]{000000} 0.99/0.99} & {\color[HTML]{000000} 0.99/0.99} & {\color[HTML]{000000} \textbf{1.00/1.00}} & {\color[HTML]{000000} 1.00/1.00} \\ \cline{2-10} 
\multirow{-4}{*}{{\color[HTML]{000000} \textbf{\begin{tabular}[c]{@{}c@{}}WADI\\ Training \\ data\end{tabular}}}} & {\color[HTML]{000000} \textbf{all(10\%)}} & {\color[HTML]{000000} 0.00/0.50} & {\color[HTML]{000000} 0.00/0.50} & {\color[HTML]{000000} 1.00/1.00} & {\color[HTML]{000000} \textbf{1.00/1.00}} & {\color[HTML]{000000} 0.99/0.99} & {\color[HTML]{000000} 0.99/0.99} & {\color[HTML]{000000} 1.00/1.00} & {\color[HTML]{000000} \textbf{1.00/1.00}} \\ \hline
\end{tabular}%
\end{table*}

The def-NN and RF models have very similar results: (1) the best score happens on testing and training data with the same noise level, indicating that the defence models perform well for similar noise to the training data; (2) both models perform best on data with `all (10\%)' and perform worst on data with `sensor (1\%)', which is reasonable as the noisier data becomes, the easier it is to be detected. We can also see the result that a model predicts better for data which is noisier than its training data. As for the difference: RF models overall perform better than def-NN, especially for WADI data. In SWaT, the model trained from `sensor (1\%)' could be generalised to others, but other models cannot be generalised. In particular, none of the models from def-NN can be generalised to all attack scenarios. To some degree, it shows that generating defences for our adversarial attacks without any information about attack details is difficult. In WADI, both models perform better than SWaT, while def-NN still has the overfitting problem when training data is noisier. These results suggest that our adversarial attacks (designed for RNN) may be able to deceive additional neural network based defences.

Overall, the defence models give us the view that neural network based defences are not as effective as statistics-based defences with respect to our adversarial attacks. Furthermore, the more information we have about the adversarial samples, the better defence model we can train. Nevertheless, from the attacker's perspective, the less noise applied, the easier it is to avoid detection. 

\begin{center}
\noindent\fbox{%
    \parbox{0.75\linewidth}{%
        \small\emph{It is hard to design a defence model without any information about adversarial samples. Even with adversarial samples as training data, a neural network defence model cannot be generalised for different systems and attack scenarios.}
    }
}
\end{center}

\subsection{Threats to Validity}

There are some limitations of the evaluation and application validity. First, though the two CPS systems are real and operational, they are still testbeds and not at the scale of industrial plants. Second, while our study shows that it is possible to degrade an anomaly detector's accuracy (while simultaneously evading rule checkers), we did not investigate whether this can be translated into an attack that breaks down the system---this is left as future work. Third, the data was collected and labelled in a manual effort, and thus there may be data points that are not accurate, and could increase the bias of the results. Finally, we only considered the recorded attacks of the datasets to train and calculate ground truth, so the methods may not work well for non-recorded attacks.

\section{Related Work}
\label{sec:related work}

In this section, we highlight some other work related to the main themes of this paper: anomaly detectors for CPSs, and adversarial attacks for CPS classifier evasion.

Anomaly detection for CPSs is a highly active research area (e.g.~\cite{Inoue-et_al17a,Das-Adepu-Zhou20a,Schmidt-Hauer-Pretschner20a,Kravchik-Shabtai18a,Cheng-Tian-Yao17a,Harada-et_al17a,Pasqualetti-Dorfler-Bullo11a,Aggarwal-et_al18a,Aoudi-et_al18a,He-et_al19a,Lin-et_al18a,Narayanan-Bobba18a,Schneider-Boettinger18a}), in which sensor and actuator data is used to identify possible anomalies in the system's processes or operation. A number of these anomaly detection schemes use knowledge of the processes' physics to apply techniques from control theory~\cite{Choi-et_al18a,Quinonez-et_al20a}, or use that knowledge to derive logical invariants over sensor and actuator states that can be monitored~\cite{Adepu-Mathur16a,Yoong-et_al21a}. Our paper, however, focuses on black box anomaly detection schemes that build models from the sensor and actuator data. These schemes can be implemented using techniques such as machine learning~\cite{Chen-Poskitt-Sun18a,Inoue-et_al17a} and data mining~\cite{Feng-et_al19a}.

Black box anomaly detection models can also be obtained through deep learning, i.e.~machine learning methods based on neural networks. Examples for industrial control systems include the deep neural network detector of Inoue et al.~\cite{Inoue-et_al17a}, the convolutional neural network detector of Kravchik et al.~\cite{Kravchik-Shabtai18a}, and the RNN of Goh et al.~\cite{goh2017anomaly}. Neural networks detectors have been evaluated on smart grids~\cite{kosek2016contextual}, gas oil plants~\cite{filonov_multivariate_2016} and a water tank~\cite{eiteneuer2018lstm}. Examples in other types of CPSs include neural networks for probabilistic estimation of brake pressure for electrified vehicles~\cite{lv2017levenberg}, an RNN for detecting anomalies in aircraft data~\cite{nanduri_anomaly_2016}, and neural networks for detecting fault data injections in vehicular CPSs with wireless communications~\cite{sargolzaei2016machine}.

Adversarial machine learning is concerned with the properties of learnt models when targeted by attackers~\cite{Huang-et_al11a}. The limitations of deep learning models in adversarial settings are well known~\cite{papernot2016limitations,moosavi2016deepfool,goodman2020advbox}, and adversarial attacks have famously been demonstrated in settings such as audio~\cite{carlini_audio_2018,carlini2016hidden} and image recognition~\cite{kurakin2016adversarial,smith2015face}, e.g.~by producing images that are completely unrecognisable by humans yet classified with 99.9\% confidence by neural networks.

Given the popularity of deep learning for anomaly detection in industrial control systems, a number of adversarial attacks have been proposed that attempt to evade these defences. Feng et al.~\cite{Feng-et_al17a} use generative adversarial networks to generate false sensor measurements that evade LSTM detectors while delivering the attack objective. Erba and Tippenhauer~\cite{Erba-Tippenhauer20a} spoof sensor values (e.g.~using precomputed patterns) and are able to evade three black box anomaly detectors published at top security conferences. Erba et al.~\cite{Erba-et_al20a} propose attacks that are able to evade reconstruction-based anomaly detectors: first, a black box approach based on autoencoders, and a white box approach based on optimisation with a detection oracle. Zizzo et al.~\cite{Zizzo-et_al19a} evaluate the impact of adversarial white box attacks based on the fast gradient sign method. Finally, Kravchik et al.~\cite{Kravchik-Biggio-Shabtai21a} fake the signals of corrupted sensors when neural network detectors are (re-)trained, poisoning the learning process so that certain cyberattacks can go undetected. One of the main differences between these works and our own is that we assume that the neural network detectors are complemented by built-in rule checkers (i.e.~invariant checkers) that must be evaded \emph{simultaneously}.

\section{Conclusion}
\label{sec:conclusion}
In this work, we presented an adversarial attack that \emph{simultaneously} evades both the RNN-based anomaly detectors and invariant-based rule checkers of real-world CPSs. Using a white-box gradient-descent approach, we craft noise to deceive detectors into assigning the wrong classification (allowing conventional attacks to be masked), then use a GA to optimise the manipulations so as to avoid violating any rule checkers that are present. We tested the defence mechanisms of two real-world critical infrastructure testbeds against our attack, finding that we were able to successfully reduce classification accuracy by over 50$\%$, suggesting that RNN-based anomaly detectors may be vulnerable against adversarial attacks even in the presence of other defence mechanisms such as rule checkers. Finally, we explored the possibility of mitigating attacks by training on adversarial samples, but found it was difficult to detect adversarial attacks in general unless they involved large amounts of noise.

\section*{Declaration of Competing Interests}

None.

\bibliography{main}

\end{document}